# Scalable $Al_2O_3$-$TiO_2$ Conductive Oxide Interfaces as Defect Reservoirs for Resistive Switching Devices

*Yang Li[1*], Wei Wang[1], Di Zhang[2], Maria Baskin[1], Aiping Chen[2], Shahar Kvatinsky[1], Eilam Yalon[1], and Lior Kornblum[1*]*

[1]The Andrew and Erna Viterbi Faculty of Electrical and Computer Engineering, Technion - Israel Institute of Technology, Haifa, 3200003, Israel

[2]Center for Integrated Nanotechnologies (CINT), Los Alamos National Laboratory, Los Alamos, New Mexico 87545, USA

**KEYWORDS**: two-dimensional electron gas (2DEG), conductive oxide interfaces, valence change memory (VCM), resistive switching, resistive switching memory, atomic layer deposition (ALD).

**ABSTRACT**

Resistive switching devices herald a transformative technology for memory and computation, offering considerable advantages in performance and energy efficiency. Here we employ a simple and scalable material system of conductive oxide interfaces and leverage their unique properties for a new type of resistive switching device. For the first time, we demonstrate an $Al_2O_3$-$TiO_2$ based valence-change resistive switching device, where the conductive oxide interface serves both as the back electrode and as a reservoir of defects for switching. The amorphous-polycrystalline $Al_2O_3$-$TiO_2$ conductive interface is obtained following the technological path of simplifying the fabrication of the two-dimensional electron gases (2DEGs), making them more scalable for practical mass integration. We combine physical analysis of the device chemistry and microstructure with comprehensive electrical analysis of its switching behavior and performance. We pinpoint the origin of the resistive switching to the conductive oxide interface, which serves as the bottom electrode and as a reservoir of oxygen vacancies. The latter plays a key role in valence-change resistive switching devices. The new device, based on scalable and complementary metal-oxide-semiconductor (CMOS) technology-compatible fabrication processes, opens new design spaces towards increased tunability and simplification of the device selection challenge.

**INTRODUCTION**

The two-dimensional electron gases (2DEGs) formed at some oxide interfaces provide a fertile ground for many physical phenomena[1–6] that do not exist in the corresponding bulk oxide material. These phenomena and the sheet resistance tunability of the 2DEG make it interesting for various device applications such as the channel in transistors[7–9] and the electrode in resistive switching memories[10–14]. Resistive switching memory, which has a metal-oxide-metal sandwiched structure, is a promising candidate for next-generation memories to address the vast data storage requirement in the big data era[15,16]. Its conductance tunability emulating the synaptic plasticity makes it applicable for brain-inspired in-memory computing[17–20], which can overcome the memory bottleneck in the traditional von Neumann architecture and drastically decrease the energy consumption.

Several 2DEGs based valence change memory (VCM) devices were reported on various materials systems[10,12–14], which share a common feature of using single-crystal $SrTiO_3$ substrates. The 2DEGs in these works function as the bottom electrodes of the devices. Originally, 2DEGs were fabricated exclusively by oxide epitaxy, typically by pulsed laser deposition (PLD) at high temperatures of 600-800 °C. The epitaxial oxide layer was subsequently replaced by amorphous oxides such as $Al_2O_3$, which lowered the deposition temperature to ~300 °C by the use of atomic layer deposition (ALD)[21]. However, the use of $SrTiO_3$ single crystal substrates hinders the scalability and makes direct integration of such devices too complicated for the CMOS process. Recent works have reported 2DEG at the interface of amorphous $Al_2O_3$ and polycrystalline $TiO_2$[8,11,22–24], both deposited in a single ALD process. The mechanism of the 2DEG formation was determined to be the formation of oxygen vacancies at the $TiO_2$ surface through a reduction caused by the Al ALD precursor in the initial stages of $Al_2O_3$ deposition.[23] These negative charges at the $TiO_2$ side of the interface result in a downward band bending at the $TiO_2$ surface, which was shown to localize the 2DEG carriers near the interface.[8] The use of low-temperature ALD opens the opportunity for scalable fabrication of resistive switching memory devices with CMOS-compatible processes and materials that can be integrated on the Si backend or a variety of other substrates. An electrochemical metallization (ECM) device of $Cu/Ti/Al_2O_3/TiO_2$ has been reported[11], where the resistive switching is based on the formation and rupture of filaments consisting of Cu metal ions from the top electrode. The 2DEG formed at the $Al_2O_3/TiO_2$ interface is inherently driven by oxygen vacancies, and it can thus act as an oxygen vacancy reservoir for VCM type resistive switching memory devices in addition to its more conventional role as a bottom electrode.

Here we report the first demonstration of VCM devices based on the 2DEGs formed at the $Al_2O_3/TiO_2$ interface. The interface is fabricated at low temperatures (<300 °C) using the scalable ALD technique. The use of the low-temperature ALD and simple binary oxide materials system provides a unique feature of the 2DEG based devices being able to be integrated with the back-end-of-line CMOS process, in contrast to most previously reported 2DEG based VCM devices[10,12–14]. Note that for the sake of convenience, we use the term 2DEG very loosely to describe conductive oxide interfaces, in many of which the conductivity is not limited to a 2D sheet exactly at the interface, and it may extend deeper into the bottom material.

Unlike standard resistive switching devices, where the bottom electrode is a metal, a distinct feature of the 2DEG electrode is that it can be switched off, via depletion by field effect[8]. This feature of 2DEGs opens prospects of depleting the device's bottom electrode via the top electrode, which would result in a strongly asymmetric behavior. The asymmetry provides the potential to design *self-rectifying* resistive switching devices[16,25]. This is an attractive approach for inhibiting sneak-path currents, which circumvents the necessity of additional transistors or selector devices in cross-bar arrays[26]. As such, self-rectifying devices can dramatically simplify the large-scale integration of resistive switching memory arrays. Our demonstration of scalable 2DEG based VCM devices paves the way towards fulfilling this promise.

**RESULTS AND DISCUSSION**

We start by understanding the 2DEGs' features and mechanism and then follow on their role in the device.

**Conductive Interface between $Al_2O_3$ and $TiO_2$**

The $Al_2O_3/TiO_2$ 2DEG sheet resistance is determined as 5000 ± 40 Ω/square using the Van der Pauw configuration. The substrate is connected in parallel to the 2DEG, but it does not significantly affect the 2DEG's role in the resistive switching given its high resistivity (≈70 kΩ·cm) and vertical distance ($SiO_2+TiO_2$ layers) from the resistive switching layer, $Al_2O_3$. The cross-sectional transmission electron microscopy (TEM) image of the $Al_2O_3/TiO_2$ heterostructure is shown in Figure 1a. The phase contrast from the $TiO_2$ layer indicates that it is primarily polycrystalline. The crystallinity of $TiO_2$ is anatase as indicated by grazing incident x-ray diffraction (GIXID, Figure 1b, ICDD card #00-21-1272), and the thickness is 14.8 ± 0.6 nm determined by X-ray reflectivity (XRR, Figure S1). No phase contrast or a clear crystal lattice is observed in the TEM image of the $Al_2O_3$ layer. Our prior results[27] and current data (Figure 1b) indicate that the $Al_2O_3$ layer is amorphous.

The mechanism behind the interface conductivity is probed by x-ray photoelectron spectroscopy (XPS) analysis of the Ti 2p region in $TiO_2$ (Figure 1c and d). A spectrum of the surface of the uncapped $TiO_2$ (also 14.8 ± 0.6 nm thick, without an $Al_2O_3$ overlayer) shows a nearly stoichiometric $Ti^{4+}$ originating in $TiO_2$ (Figure 1c). A similar sample with a 2 nm thick $Al_2O_3$ capping layer shows a significant $Ti^{3+}$ component, constituting 15 ± 0.5% of the Ti 2p total peak area (Figure 1d). This lower oxidation state results from a reduction of the $TiO_2$ by the Al precursor during the early stages of the ALD-$Al_2O_3$ process[23], leading to the generation of oxygen vacancies. These oxygen vacancies act as n-type dopants in the otherwise insulating $TiO_2$, and they are therefore the source of the 2DEG[8]. This XPS fingerprint of 2DEG formation is also commonly observed in 2DEGs based on $SrTiO_3$ substrates[21,28–30]. Samples without the $Al_2O_3$ layer, showing no $Ti^{3+}$ features (Figure 1c), are electrically insulating ($R_s>10^9$ Ω/square, out of test equipment limit), showing only negligible contribution of conductivity from the Si substrate below (or no measurable conductivity when deposited on a glass substrate for comparison).

Therefore, the process described here for $Al_2O_3$ deposition results in the formation of 2DEGs at the $Al_2O_3/TiO_2$ interface, as observed by the interface conductivity and the significant $Ti^{3+}$ component (Figure 1d).

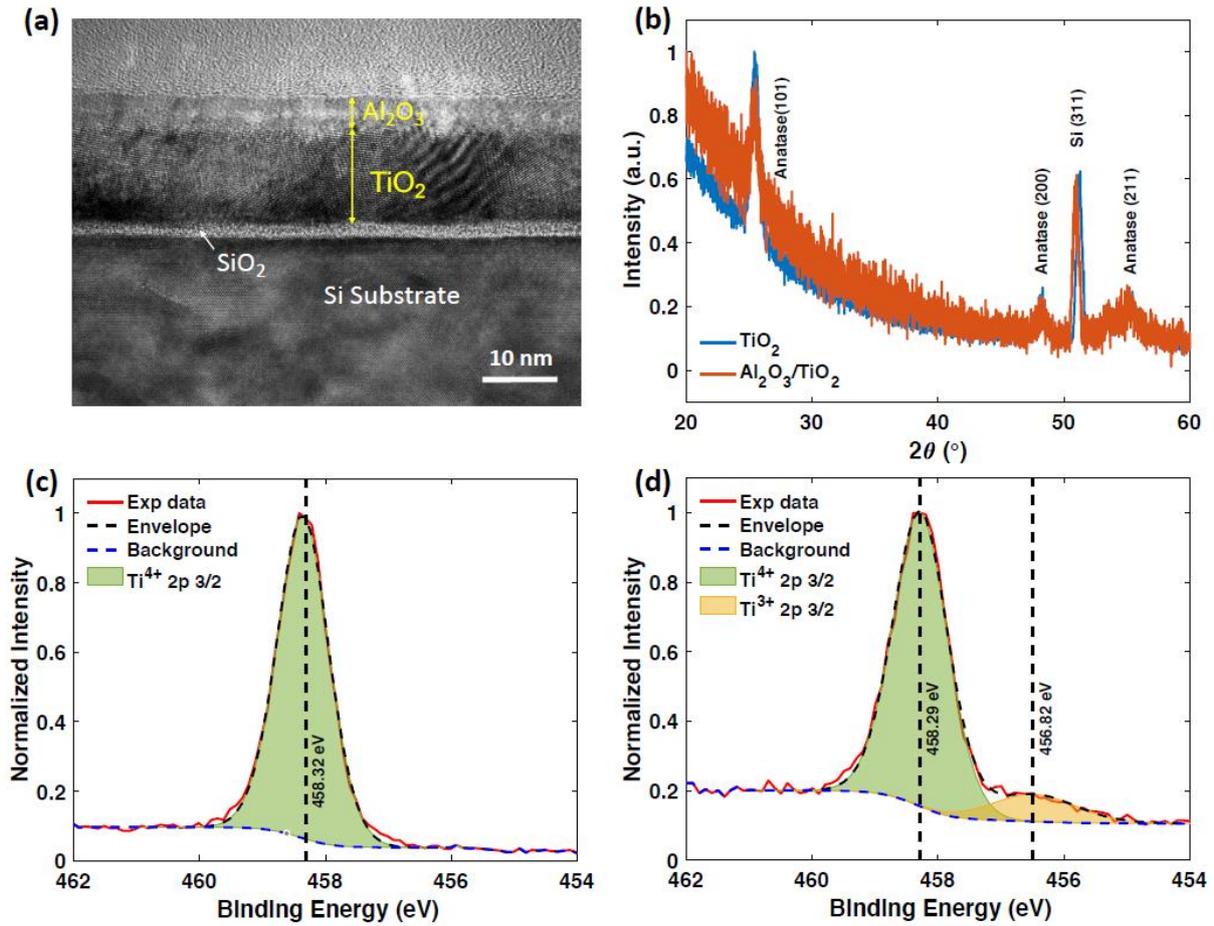

Figure 1. The uncapped TiO$_2$ and the Al$_2$O$_3$/TiO$_2$ 2DEG heterostructure material characterization. (a) TEM cross-sectional view of the Al$_2$O$_3$/TiO$_2$ film stack. (b) Grazing incident X-ray diffraction (GIXRD) of the uncapped TiO$_2$ film and the Al$_2$O$_3$/TiO$_2$ 2DEG film stack on SiO$_2$/Si substrates. The peaks are at 2θ=25.5°, 48.2°, 55.0° correspond to anatase (101), (200), and (211), respectively. The peak at 2θ=51.0° corresponds to Si (311). X-ray photoelectron spectroscopy (XPS) spectra of Ti 2p of (c) the uncapped TiO$_2$ film and (d) the Al$_2$O$_3$/TiO$_2$ 2DEG heterostructure. The Al$_2$O$_3$ layer is 2nm thick.

Here the Al$_2$O$_3$ (2nm) acts as a protection layer for the 2DEG formed in the surface of TiO$_2$[29]. It was intentionally made thin to allow the XPS to probe the underlying TiO$_2$, whereas, in the device to follow, thicker Al$_2$O$_3$ layers (5 nm) will be employed. The reduction process may well extend the range of the oxygen vacancies into the TiO$_2$ layer, making the conductivity not limited strictly to the Al$_2$O$_3$/TiO$_2$ interface. Therefore, the use of the term 2DEG is not rigorously justified, rather it is used here loosely for convenience [31].

**2DEG Based VCM Devices**

Having set the ground with the analysis of the 2DEGs, we then fabricate VCM devices from Pt (50 nm)/Al$_2$O$_3$ (5 nm)/TiO$_2$ (14.8 nm) structures. The Pt layer acts as the top electrode, and the Al$_2$O$_3$ layer acts as the resistive switching layer. The conductive 2DEG at the Al$_2$O$_3$/TiO$_2$ interface acts as the bottom electrode and as the oxygen vacancies reservoir for the VCM device here. The 2DEG electrode is contacted from contacts at the corners of the sample (Figure 2a).

When operating the device, we apply the voltage on the Pt top electrode and ground the 2DEG bottom electrode (Figure 2a). The *I-V* curves of the forming cycle and a typical switching cycle are shown in Figure 2b. Five different devices are tested under DC mode using voltage sweeps, and each device is switched for 20 cycles after the forming process (Figure 2c), exhibiting comparable behavior and low device-to-device variability. The detailed switching parameters are provided in Figure S2.

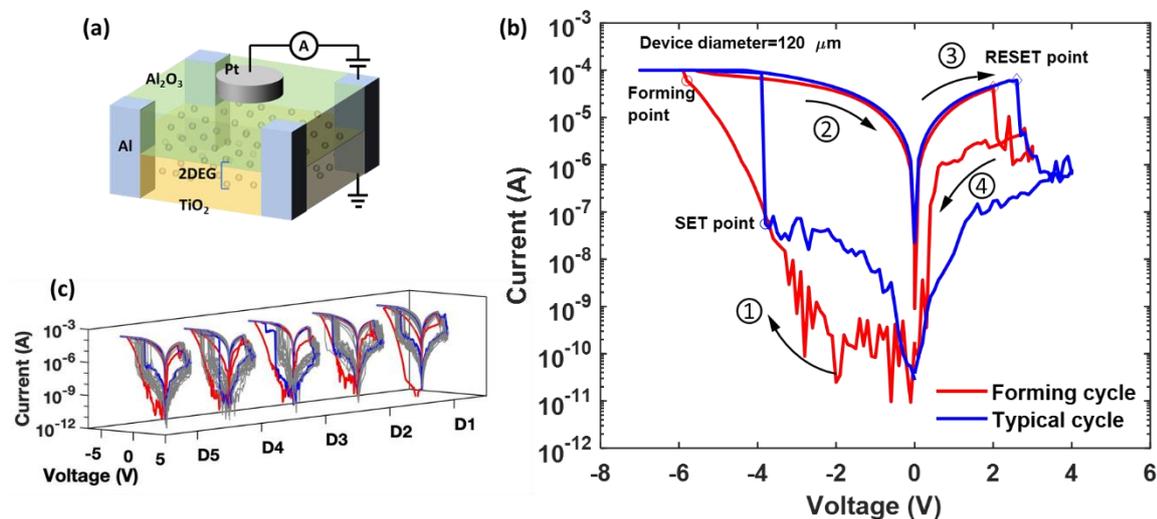

Figure 2. The Pt/Al$_2$O$_3$/TiO$_2$ device structure and DC mode resistive switching *I-V* curves. (a) Schematic illustration of the device structure with the 2DEG acting as the bottom electrode. The Pt acts as the top electrode and the Al$_2$O$_3$ acts as the resistive switching layer. The voltage is always applied to the Pt top electrode, and the 2DEG is always grounded during the tests. (b) *I-V* curves of a forming cycle and a typical switching cycle in DC mode. The forming, SET, and RESET points are noted on the curves. The annotations of ① ② ③ ④ show the sequence of the voltage sweep and the arrows show the voltage sweep direction. (c) *I-V* curves of five devices. The red lines are the forming cycles. The blue lines represent the typical resistive switching cycles. The grey lines correspond to all *I-V* cycles after the forming cycles.

In addition to the 2DEG's role as the back electrode, it further functions as the source of oxygen vacancies responsible for the resistive switching behavior of the Al$_2$O$_3$ layer. In a previous work, we deposited Al$_2$O$_3$ layers with an identical recipe on conductive Nb:SrTiO$_3$ layers, where no interface vacancies were formed. These structures exhibit robust insulating properties, and no hysteresis behavior.[32] The key difference from the current case is the reduction of the TiO$_2$ surface, allowing us to pinpoint the interface as the source of vacancies for resistive switching, which are injected (forming) into the $Al_2O_3$ layer.[13] This differs from typical memristors where the resistive switching layer is intentionally rich with defects, whereas our approach allows the use of an initially low-defect and insulating $Al_2O_3$, which heralds a memory window of ~4 orders of magnitude at ±0.2 V.

The retention of high resistive state (HRS) and low resistive state (LRS) using compliance current ($I_{CC}$) of $10^{-4}$ A (using the current limit function of the testing instrument) are tested at room temperature using 0.1 V read voltage and both resistance states remain stable for $10^4$ s without degradation (Figure S3), which is comparable to other 2DEG based VCM devices on $SrTiO_3$ substrates[10,12,14].

In addition to DC switching analysis, the endurance of the devices is tested under pulsed switching using 100 μs width voltage pulses (Figure 3). The resistance window is set to be 10, with the LRS lower than $10^6$ Ω and the HRS higher than $10^7$ Ω (Figure 3a). An operate-and-verify programming scheme[33–36] is used here, which means more than a single pulse may be used in the SET or RESET operations. The voltage height in both SET and RESET will increase until the resistance is tuned to the destination resistance value. The details of the operate-and-verify programming scheme are shown in Figure S4-S6. The reason for using the operate-and-verify programming scheme is that the device might not reach the desired resistance range after a single pulse operation. To ensure the device's resistance is indeed programmed to the desired value, a verification step is necessary. This operate-and-verify programming scheme requires higher energy consumption compared to single-pulse programming scheme, but it is more reliable and it guarantees the resistance modulation success for each cycle. All the HRS and LRS in both DC mode and pulse mode are read at a voltage of 0.1 V. Due to the intrinsic stochastic nature of the devices' switching process which includes the formation and rupture of the conductive filament, the devices exhibit cycle-to-cycle variation (Figure 3a). The LRS values vary from 77 kΩ to 1 MΩ and follow a normal distribution (μ=7.2×$10^5$ Ω, σ=2.1×$10^5$ Ω). The HRS vary from 10 MΩ to 87 GΩ and follow a lognormal distribution (Figure 3b). The resistances of the device show no observable degradation after $10^3$ cycles of operation.

About 98% of the RESET cycles require no more than 2 pulses to switch the device from HRS to LRS, and about 85% of the SET cycles require no more than 3 pulses to switching the device from LRS to HRS (Figure 3c and 3d). The SET voltage and the number of pulses required to switch the device from HRS to LRS increase as the cycle number increases (Figure 3c), indicating that further engineering will be necessary to mature these devices for practical applications.

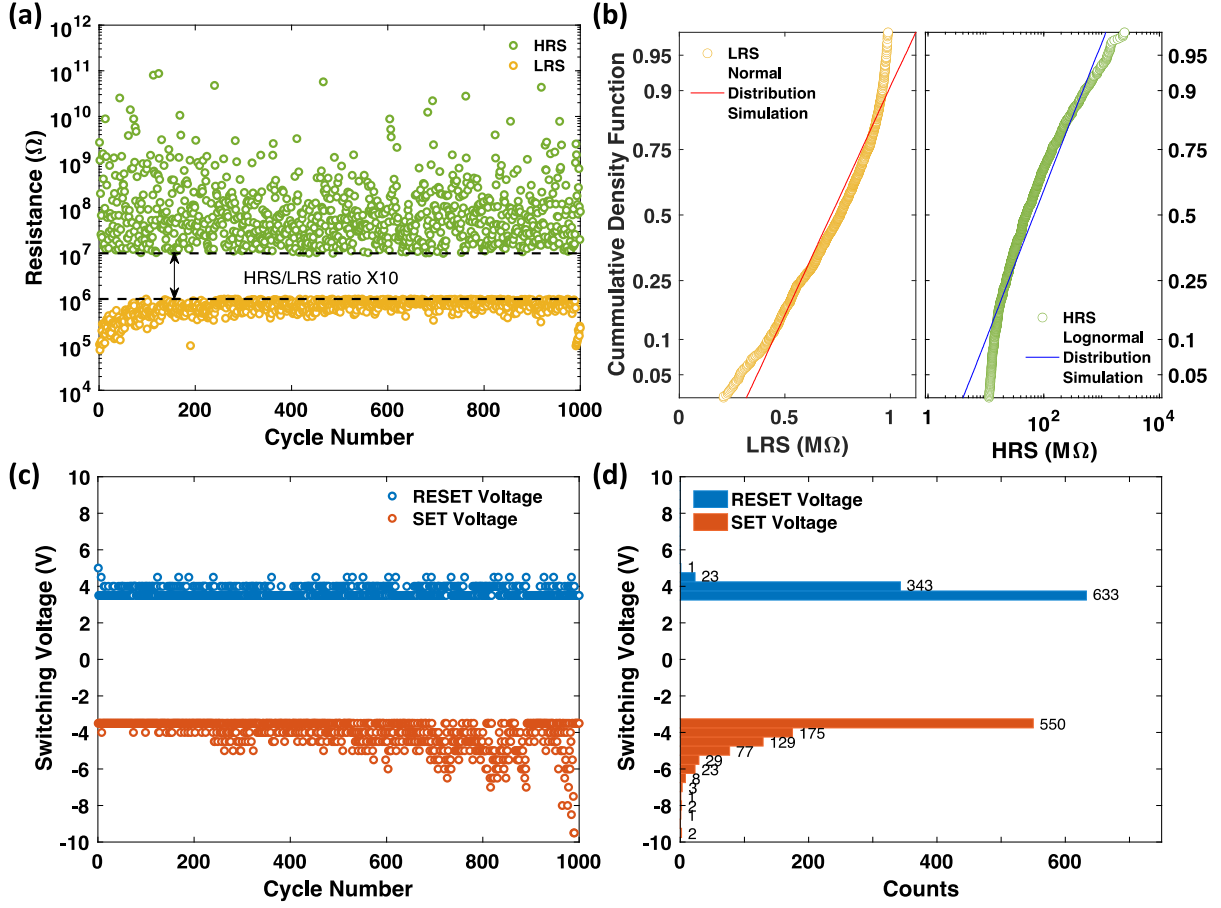

Figure 3. Endurance analysis of Pt/Al$_2$O$_3$/TiO$_2$ devices under pulse mode using 100 μs voltage pulses. (a) The resistances of 10$^3$ cycles under pulse operation mode with a ten times resistance window defined in the program. (b) The cumulative density function of the resistances. The red line is the simulated normal distribution of the LRS ~ normal ($\mu_1$=7.2×10$^5$, $\sigma_1$=2.1×10$^5$). The blue line is the simulated lognormal distribution of the HRS ~ lognormal ($\mu_2$=16.55, $\sigma_2$=2.11). (c) The last SET and RESET voltage pulse height in each cycle of 10$^3$ resistive switching cycles. (d) The histogram of the switching voltages with the count numbers labeled right to the bars.

**Switching Mechanism**

The resistive switching behavior of the Pt/Al$_2$O$_3$/TiO$_2$ VCM devices can be explained by the formation and partial rupture of the conductive filaments (CFs) consisting of oxygen vacancies[20,37–40] inside the Al$_2$O$_3$ layer, as illustrated in Figure 4.

The pristine device starts with no CFs inside the Al$_2$O$_3$ layer (Figure 4a). During the forming process, an external negative voltage is applied to the Pt electrode and the positively-charged oxygen vacancies migrate towards the top electrode which leads to the formation of CF consisting of oxygen vacancies (Figure 4b→Figure 4c). The device switches from initial high resistance state to low resistance state.

In the RESET process, a positive voltage is applied to the Pt top electrode and the bottom electrode is grounded. The oxygen vacancies migrate towards the bottom electrode and the gap length Δ between the remnant CF grows, which leads to the rupture of CF (Figure 4c → Figure 4d → Figure 4e). The device resistance switches from low resistance state to high resistance state.

In the SET process, a negative voltage is applied to the top electrode and the bottom electrode is grounded. At first, the length of the gap Δ decreases until the filament forms a continuous filament connecting the top and the bottom electrode (Figure 4e → Figure 4f). Only after the continuous CF forms, the diameter of the CF begins to increase until the SET process finishes (Figure 4f → Figure 4g). The device resistance switches from high resistance state to low resistance state.[41]

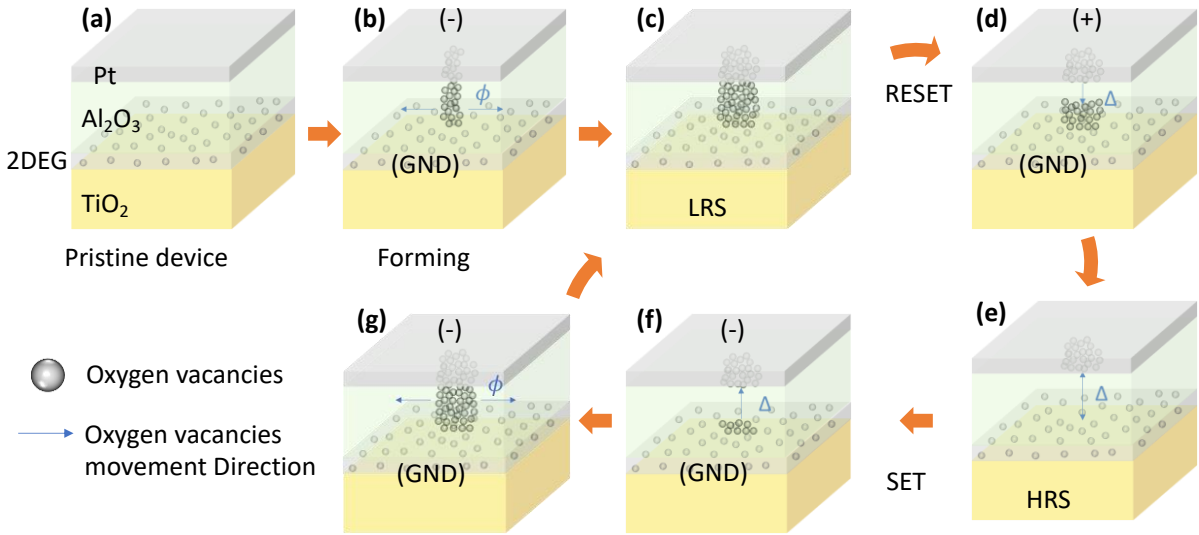

Figure 4. Schematic illustration of the filamentary resistive switching process of Pt/Al$_2$O$_3$/TiO$_2$ VCM devices. (a) The initial state of a pristine device. (b) The forming process: formation of a continuous CF consisting of oxygen vacancies connecting the top and the bottom electrode, and the diameter of the CF increases until the forming voltage sweep finishes. (c) The low resistance state (LRS) of the device. (d) The RESET process: partial rupture of the conductive filament with the external positive voltage applied to the Pt top electrode and the oxygen vacancies drifting towards the 2DEG electrode, and the gap Δ between the remnant CFs increases. (e) The high resistance state (HRS) of the device. (f) and (g) SET process: the gap Δ between the remnant CFs decreases first, and the CF diameter ϕ increases only after a continuous CF is formed.

One issue of using 2DEG as the bottom electrode is that the resistance of the 2DEG needs to be further decreased for large crossbar array device integrations. As for the advantages, we highlight the prospects of using 2DEGs as a bottom electrode that can be depleted, thus opening simple routes towards self-rectifying devices. Realizing the potential for self-rectification here requires controlling of the depletion behavior of the 2DEG, which can be obtained by tuning of the gate effective work function[42], engineering (or adding) the insulator layer[8], and with designing the device structure and contacts with an emphasis on the series resistance.

In addition, if the inert top electrode is replaced by an oxide conductor, e.g., indium tin oxide (ITO)[22], it can enable full oxide implementation of the resistive switching devices. This will pave the way toward transparent memory applications. This is another potential benefit of using binary oxide-based 2DEG as the bottom electrode and the oxygen reservoir in resistive switching devices.

**CONCLUSIONS**

VCM devices based on 2DEG on Si are demonstrated for the first time. The 2DEG acts as both the bottom electrode and the oxygen reservoir. The Pt/Al$_2$O$_3$/TiO$_2$ VCM devices show binary resistive switching in DC and pulse modes. An endurance of $10^3$ cycles and retention of $10^4$ s are obtained. The devices switch between HRS and LRS due to the formation and partial rupture of conductive filaments consisting of oxygen vacancies inside the Al$_2$O$_3$ layer. The use of the Si substrate and simple binary oxide materials of Al$_2$O$_3$ and TiO$_2$ demonstrate the feasibility of integrating 2DEG into Si-based and CMOS-compatible devices. The ability to deplete the 2DEGs provides a route towards self-rectifying devices, and the possibility of replacing the top metal electrode with a transparent oxide electrode also allows the device to be extended to wearable and flexible applications.

**EXPERIMENTAL SECTION**

The substrates used are (100) n-type unintentionally-doped Si (MTI Corp., resistivity ≈ 70k Ω·cm) with ~2 nm native SiO$_2$. The TiO$_2$ films are deposited on the substrates by atomic layer deposition (ALD, Ultratech/Cambridge Nanotech Fiji G2) at 250 °C using tetrakis(dimethyl amido)titanium (TDMAT) and Ar:O$_2$ (4:1) plasma as the Ti and oxygen precursors, respectively. The TDMAT is heated to 75 °C in the bubbler and introduced to the ALD reactor by 30 sccm of Ar carrier gas. The TiO$_2$ deposition sequence consists of TDMAT injection (0.025 s)/purge (10 s)/Ar:O$_2$ plasma (5 s)/purge (5 s). The samples are kept inside the ALD chamber under vacuum (~2×10$^{-2}$ Torr) after the TiO$_2$ deposition. The temperature is increased from 250 °C to 300 °C and stabilized at 300 °C, taking 15 minutes altogether. This step was found to be important for reducing the sheet resistance. Subsequently, 10 trimethylaluminum (TMA) reducing pulses (0.1 s pulse duration) are injected into the chamber prior to Al$_2$O$_3$ deposition. Finally, 2 nm (for XPS characterization) or 5 nm (for device testing) Al$_2$O$_3$ thin films are deposited on the TiO$_2$ at 300 °C. TMA and H$_2$O are used as the Al and the oxygen precursors, respectively. The Al$_2$O$_3$ deposition sequence[32] consists of TMA injection (0.1

s)/purge (10 s)/H$_2$O (0.3 s)/purge (5 s). Pt top electrodes (50 nm thick) are deposited on top of the Al$_2$O$_3$ layer through a shadow mask with a diameter of 120 μm using e-beam deposition. Contacts to the 2DEG are done by scratching the samples' corners from the top of the samples' surface following depositing a layer of 50 nm thick Al by e-beam deposition (Evatec BAK-501A) at room temperature on the surface of the samples' edges.

Cross-sectional specimens were prepared through a conventional TEM sample preparation routine. Starting with cutting and gluing, the TEM specimens are then ground, planar parallel polished, and further thinned in the center by dimpling. Ar ion milling is used to obtain a perforation and electron transparent thin area with a Gatan Precision Ion Polishing System (PIPS II 695, Gatan Inc.). The microstructure is characterized by FEI Tecnai F30 analytical TEM operating at 300 kV.

The films' crystallinity is analyzed by grazing incident X-ray diffraction (GIXRD), and the film's thickness is analyzed by X-ray reflectivity (XRR) using Rigaku SmartLab diffractometer. The GIXRD incident angle is 0.5°. The samples are mounted on zero-diffraction discs (MTI Corp.) during the test.

X-ray Photoelectron Spectroscopy (XPS) measurements are performed in UHV (2.5×10$^{-10}$ Torr base pressure) using the 5600 Multi-Technique System (PHI, USA). The samples are irradiated with an Al Kα monochromated source (1486.6 eV), and the outcome electrons are analyzed by a Spherical Capacitor Analyzer using a slit aperture of 0.8 mm. All spectra are shifted to align the adventitious C1s peak at 284.8 eV.

The sheet resistance of the 2DEG is measured using the Van der Pauw configuration at room temperature. The sheet resistance and DC device analysis is done using a light-sealed probe station with a Keithley 2450 source measurement unit. For resistive switching measurements the current compliance is applied through the current limit function of the instrument. AC pulse testes are measured using an Agilent B1500A.

**CORRESPONDING AUTHORS**

*E-mail: yangli@campus.technion.ac.il

*E-mail: liork@ee.technion.ac.il

ACKNOWLEDGEMENTS

The authors are grateful for the support of the Israeli Science Foundation (ISF Grant No. 375/17). Partial support in the fabrication and characterization of the samples was provided by the Technion's Micro-Nano Fabrication & Printing Unit (MNF&PU) and the Russell Berrie Nanotechnology Institute (RBNI). The authors would like to thank the assistance and support from Dr. Kamira Weinfeld and Dr. Pini Shekhter for XPS characterization, Mrs. Valentina Korchnoy for ALD, Mr. Arkadi Gavrilov for e-beam evaporation, and Dr. Maria Koifman Khristosov for XRD characterization. Y. Li acknowledges partial support from a Technion Fellowship. W. Wang was supported in part at the Technion by the Aly Kaufman Fellowship.

The work at Los Alamos National Laboratory was supported by the NNSA's Laboratory Directed Research and Development Program, and was performed, in part, at the CINT, an Office of Science User Facility operated for the U.S. Department of Energy Office of Science. Los Alamos National Laboratory, an affirmative action-equal opportunity employer, is managed by Triad National Security, LLC for the U.S. Department of Energy's NNSA, under contract 89233218CNA000001.

# Supplementary Information for:

# Scalable Al$_2$O$_3$-TiO$_2$ Conductive Oxide Interfaces as Defect Reservoirs for Resistive Switching Devices

*Yang Li[1*], Wei Wang[1], Di Zhang[2], Maria Baskin[1], Aiping Chen[2], Shahar Kvatinsky[1], Eilam Yalon[1], and Lior Kornblum[1*]*

[1]The Andrew and Erna Viterbi Faculty of Electrical and Computer Engineering, Technion - Israel Institute of Technology, Haifa, 3200003, Israel

[2]Center for Integrated Nanotechnologies (CINT), Los Alamos National Laboratory, Los Alamos, New Mexico 87545, USA

The TiO$_2$ film thicknesses are determined by using x-ray reflectometry (XRR, Figure S1).

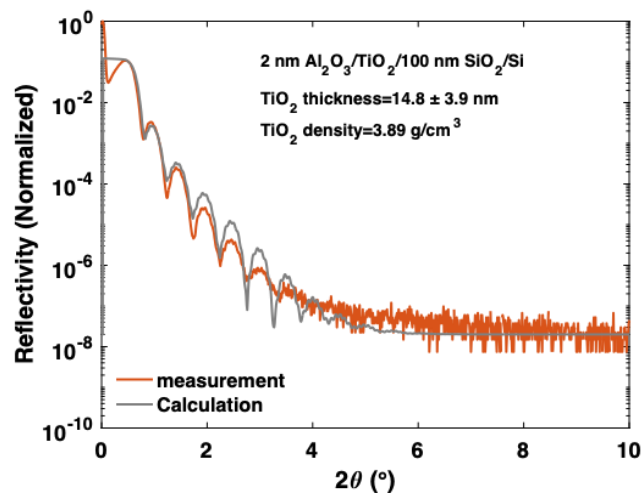

**Figure S1**. X-ray reflectometry (XRR) of the 2 nm Al$_2$O$_3$/TiO$_2$/native SiO$_2$/Si film stacks. The thickness of the TiO$_2$ was calculated to be 14.8 ± 0.6 nm using the TiO$_2$ film density of 3.89 g/cm$^3$.

The XPS spectra are fitted using the CasaXPS software with a Shirley background and a Lorentzian-Gaussian line shape with a ratio of 30%. During the XPS data analysis, the Ti 2p 3/2 peak and Ti 2p 1/2 peak area ratio is fixed to 0.5. The XPS spectra are modeled using the parameters listed in Table S 1. The peak positions and full width at half maximum (FWHM) are similar to the reference[39].

*Table S 1 The binding energies and full width at half maximum (FWHM) of Ti oxidation states.*

| Oxide State | Binding Energy (eV) | | | | | |
|---|---|---|---|---|---|---|
| | Uncapped TiO$_2$ | | TiO$_2$/Al$_2$O$_3$ | | D. Gonbeau et al.[39] | |
| | Peak position | FWHM | Peak position | FWHM | Peak position | FWHM |
| Ti$^{4+}$ 2p 3/2 | 458.32 eV | 0.95 eV | 458.29 eV | 1.04 eV | 458.7 eV | 1.5 eV |
| Ti$^{3+}$ 2p 3/2 | -- | -- | 456.82 eV | 1.88 eV | 456.6 eV | 1.4 eV |

### DC and AC resistive switching mode

A forming process, which is usually conducted in DC mode, is required to trigger the resistive switching behavior of the device. In the forming process, the voltage sweeps forward from 0 V to -6 V (red line annotated by ① in Figure 2b) and backward from -6 V to 0 V (red line annotated by ② in Figure 2b), i.e., the double voltage sweep between 0 V and -6 V, with a compliance current of $10^{-4}$ A. The forming process switches the device resistance from the initial high resistance state (HRS) to low resistance state (LRS) and leads to the formation of conductive filaments (CFs) that are consist of oxygen vacancies[40] inside the $Al_2O_3$ layer.

Except the first resistance switching from HRS to LRS is called forming, other following processes of switching the device resistance from HRS to LRS are called SET. In DC mode, the SET process is operated using a dual voltage sweep between 0 and -7 V with a compliance current of $10^{-4}$ A. In pulse mode, the SET process is completed by applying negative voltage pulses with 100 μs pulse width. The height of the voltage pulse increases from -3.5 V to -9.5 V with a step size of 0.5 V until the device is switched back to LRS (Figure S 4).

The opposite process of witching the device resistance from LRS back to HRS by partial rupturing of the CFs is called RESET. In DC mode, the RESET process uses a dual voltage sweep between 0 V and 3 V after the forming process (red line annotated by ③ and ④ in Figure 2b) or uses a dual voltage sweep between 0 V and 4V after the SET process. In pulse mode, the RESET process uses positive voltage pulses with 100 μs width. A similar operate-and-verify programming scheme[28,29] is used (Figure S 5).

A full round of switching the device resistance from HRS to LRS and then back to HRS is called a switching cycle. The blue lines in Figure 2b represent a typical *I-V* curve of the resistive switching cycle after the first forming-and-RESET cycle. Figure S 6 shows a complete cycle of the switching process in pulse mode.

The switching points are the points where the current jumps up or down abruptly, indicating an abrupt change in device resistance. The switching points can be extracted from the DC *I-V* curves, as noted in Figure 2b. The voltages and the currents of the switching points are defined as the switching voltages and the switching currents, respectively. All the resistances are read at a voltage of 0.1 V.

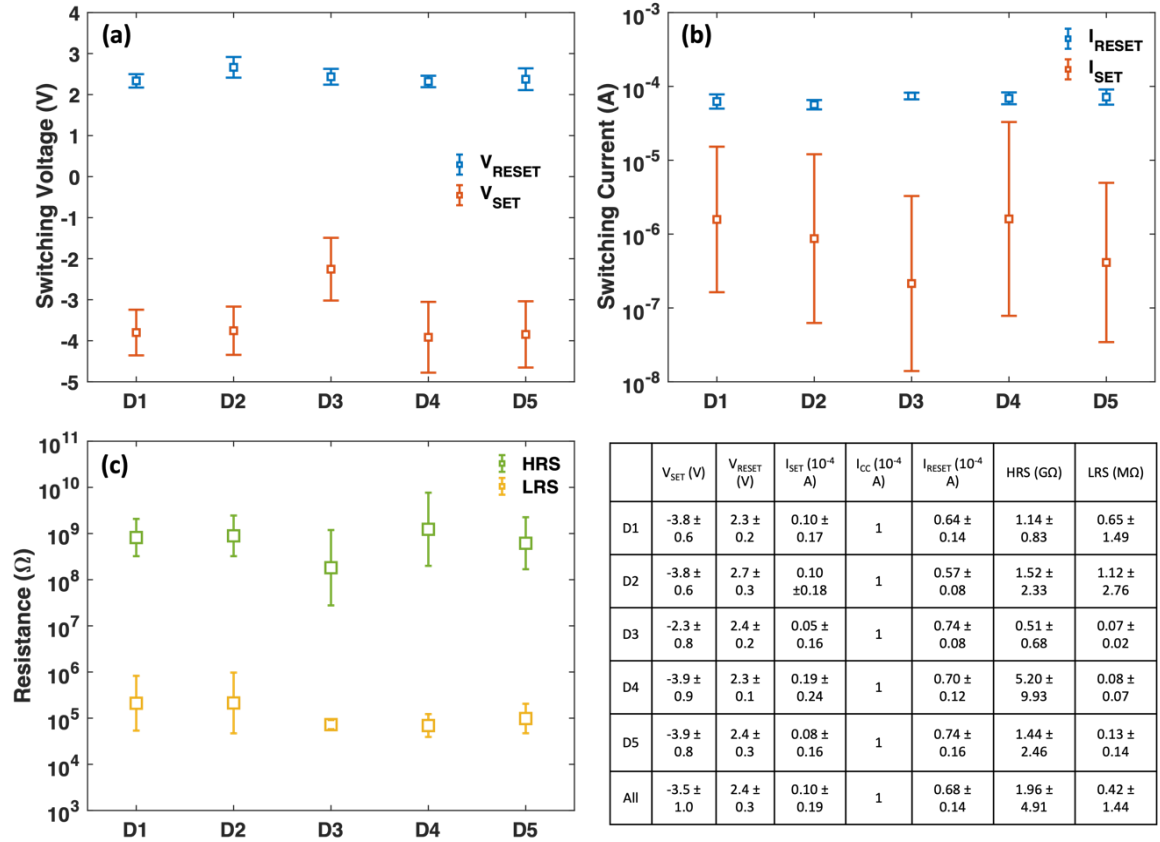

**Figure S2.** Switching parameters of five different Pt/Al$_2$O$_3$/TiO$_2$ devices under DC voltage sweep mode. (a) Error bar graph of the SET and RESET switching voltages, (b) error bar graph of the SET and RESET switching currents, (c) error bar graph of the HRS and LRS. The table on the bottom-right side lists all the parameters in detail.

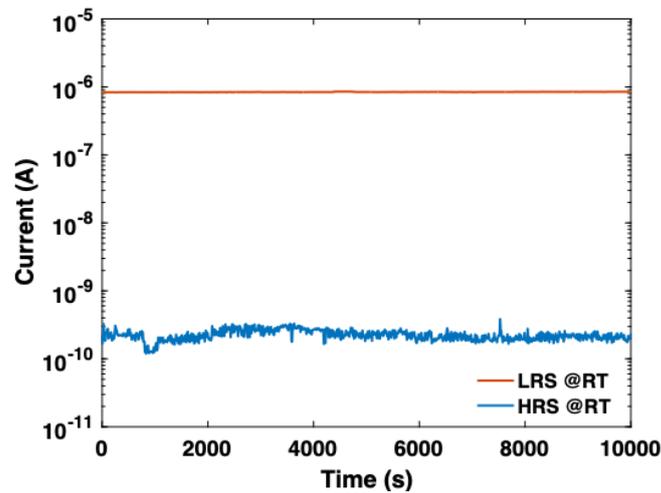

**Figure S3.** Retention time of HRS and LRS at room temperature.

An operate-and-verify programming scheme is used here, which means more than a single pulse may be used in the SET or RESET operations. The voltage height in both SET and RESET process will increase until the resistance is tuned to the destination resistance value. The details of the operate-and-verify programming scheme is shown in Figure S4-S6. The reason for using the operate-and-verify programming scheme is that the device might not reach the desired resistance range after a single pulse operation. To ensure the device's resistance is indeed programmed to the desired value, the verification step is necessary. This operate-and-verify programming scheme requires higher energy consumption compared to single-pulse programming scheme, but is more reliable and guarantees the resistance modulation success for each cycle.

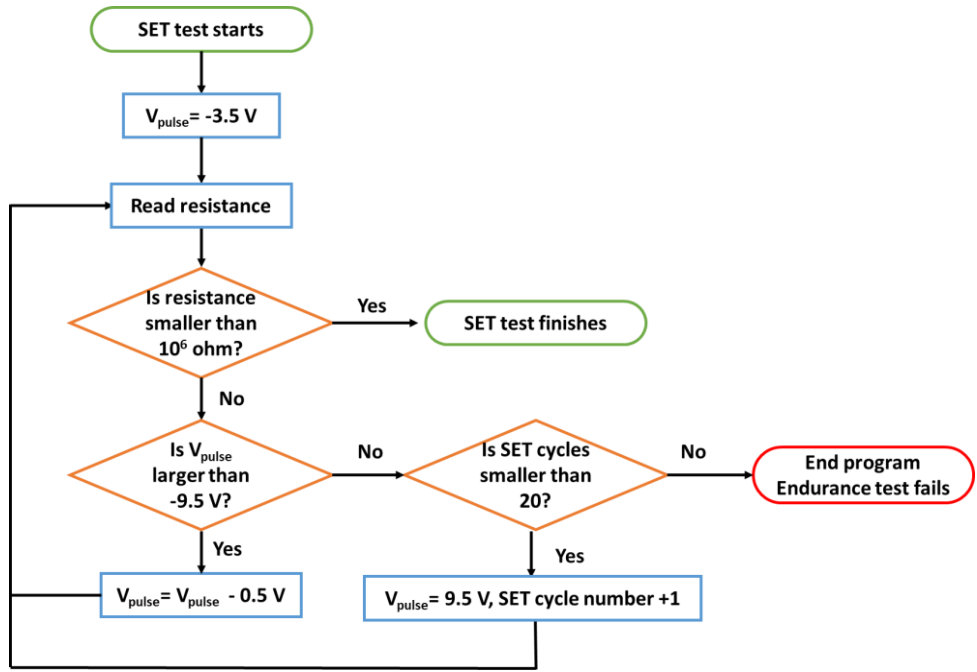

**Figure S4.** Chart flow of the SET process under pulse mode.

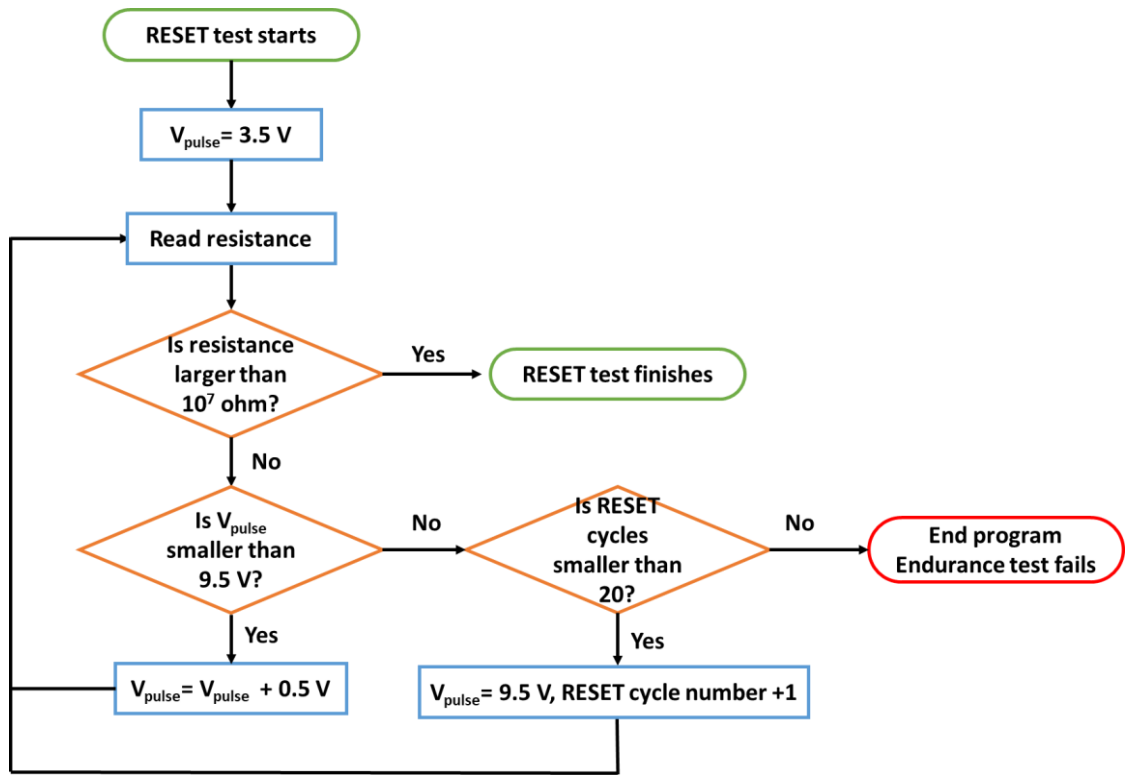

**Figure S5**. Chart flow of the RESET process under pulse mode.

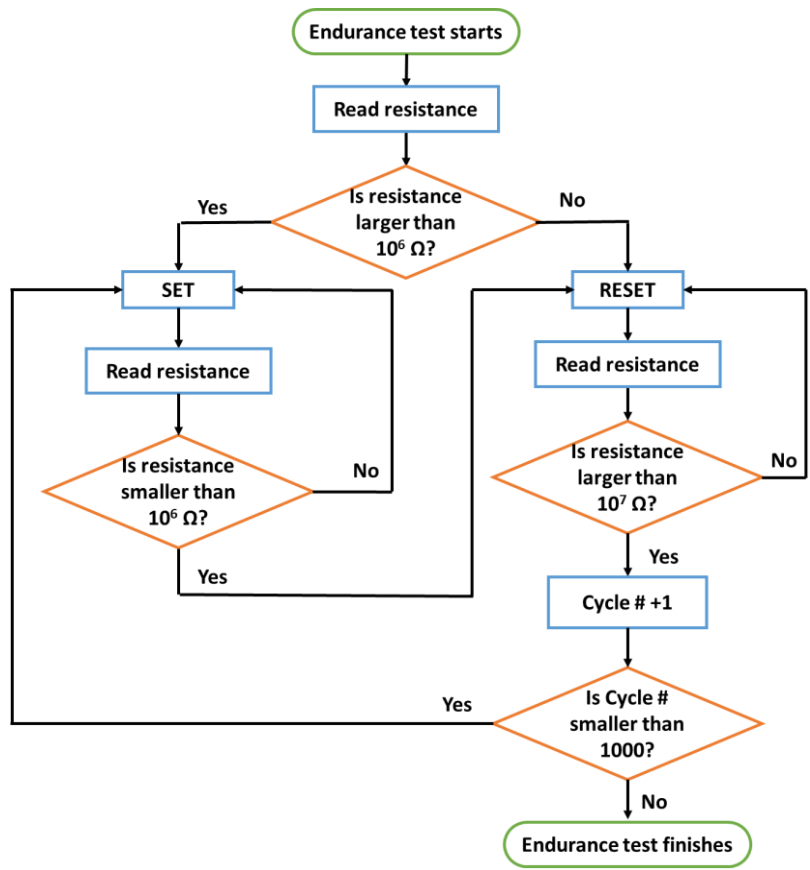

**Figure S6.** Chart flow of the endurance test under pulse mode.